\documentclass[aps, pra, superscriptaddress, reprint]{revtex4-2}

\usepackage{hyperref}
\usepackage{physics}
\usepackage{xcolor}
\usepackage{amssymb}
\usepackage{mathtools}
\usepackage{bbm}
\usepackage{mathrsfs}
\usepackage{verbatim}

\begin{document}


\title{
    Steady-state entanglement scaling in open quantum systems:\\
    A comparison between several master equations
}

\date{\today}

\author{Antonio D'Abbruzzo}
\email[Corresponding author: ]{antonio.dabbruzzo@sns.it}
\affiliation{Scuola Normale Superiore, 56126 Pisa, Italy}

\author{Davide Rossini}
\affiliation{Dipartimento di Fisica dell'Università di Pisa and INFN, Sezione di Pisa, I-56127 Pisa, Italy}

\author{Vittorio Giovannetti}
\affiliation{Scuola Normale Superiore, NEST, and Istituto Nanoscienze-CNR, 56126 Pisa, Italy}

\author{Vincenzo Alba}
\affiliation{Dipartimento di Fisica dell'Università di Pisa and INFN, Sezione di Pisa, I-56127 Pisa, Italy}

\begin{abstract}
    We investigate the scaling of the fermionic logarithmic negativity (FLN) between complementary intervals in the steady state of a driven-dissipative tight-binding critical chain, coupled to two thermal reservoirs at its edges.
    We compare the predictions of three different master equations, namely a nonlocal Lindblad equation, the Redfield equation, and the recently proposed universal Lindblad equation (ULE).
    Within the nonlocal Lindblad equation approach, the FLN grows logarithmically with the subsystem size $\ell$, for any value of the system-bath coupling and of the bath parameters.
    This is consistent with the logarithmic scaling of the mutual information analytically demonstrated in [\href{https://link.aps.org/doi/10.1103/PhysRevB.106.235149}{Phys. Rev. B \textbf{106}, 235149 (2022)}].
    In the ultraweak-coupling regime, the Redfield equation and the ULE exhibit the same logarithmic increase; such behavior holds even when moving to moderately weak coupling and intermediate values of $\ell$.
    However, when venturing beyond this regime, the FLN crosses over to superlogarithmic scaling for both equations.
\end{abstract}

\maketitle


\section{Introduction} \label{sec:intro}

Understanding how genuine quantum properties of out-of-equilibrium many-body systems are affected by an external environment~\cite{Breuer2002,Rivas2011} is of paramount importance from a fundamental point of view and is also relevant for a number of modern applications, such as quantum computation~\cite{Verstraete2009}, quantum thermodynamics~\cite{Binder2019}, and quantum chemistry~\cite{Cao2019}.
Here we are interested in understanding how entanglement~\cite{Amico2008,Calabrese2009b,Eisert2010,Laflorencie2016}, which is the distinctive feature of quantum mechanics, and its large-scale behavior are affected by unavoidable external disturbances~\cite{Aolita2015}.

The scaling of entanglement has been investigated extensively in out-of-equilibrium \textit{closed} quantum many-body systems, where one typically observes a robust growth of entanglement when increasing the size.
For example, in integrable systems the entanglement dynamics can be described using a hydrodynamic picture in terms of well-defined quasiparticles~\cite{Calabrese2005,Fagotti2008,Alba2017,Alba2018,Alba2021b,Klobas2021,Bertini2022}: in this case, the steady state exhibits volume-law entanglement entropy, which reflects the extensive character of the thermodynamic entropy.
This is accompanied by an area-law scaling for the mutual information and the logarithmic negativity~\cite{Coser2014,Alba2018,Alba2019}, which is consistent with the expected area-law mutual information that is found in thermal states~\cite{Wolf2008}.
Mild area-law violations, such as logarithmic ones, have been observed in steady states arising after quantum quenches in one-dimensional systems (see Ref.~\cite{Maric2023} for a recent result).

In \textit{open} quantum many-body systems it is more challenging to quantify entanglement.
First, since the global state of the system is mixed, neither the von Neumann entropy nor the mutual information are proper measures of entanglement~\cite{Groisman2005}.
In this situation, one can employ the logarithmic negativity~\cite{Vidal2002,Plenio2005,Eisert2001}.
However, a general technique for handling the dynamics of open quantum systems is still missing and one has to resort to approximations, which typically involve a weak-coupling assumption that leads to the so-called Lindblad equation~\cite{Breuer2002}.
The hydrodynamic description of entanglement-related quantities has been applied to the Lindblad framework, at least for quadratic master equations in the presence of global dissipation~\cite{Alba2021a,Carollo2022,Alba2022b,Murciano2023,Caceffo2024}.
Very recently, these results have been extended to describe the dynamics of the logarithmic negativity~\cite{Alba2023,Caceffo2024b}.
Still, in the presence of global system-environment interaction, the steady-state entanglement content is modest, since the negativity exhibits an exponential approach with time to a steady-state area law.
The scenario is somewhat different in the presence of localized dissipation~\cite{Alba2022c,Alba2022a,Caceffo2023,Fraenkel2021,Fraenkel2024}, which can act as sources of entanglement.

Recently, in Ref.~\cite{DAbbruzzo2022} it was shown that boundary-driven free-fermion chains~\cite{Landi2022} that give rise to current-carrying steady states can exhibit area-law violations of entanglement (see also Ref.~\cite{Ribeiro2017} for a similar result).
Specifically, it was shown analytically that the steady-state mutual information between two complementary regions exhibits logarithmic growth with the size, as long as the bulk of the system is tuned to the ground-state critical point.
This resembles the analogue ground-state entanglement scaling, even though with a nonuniversal prefactor of the logarithmic growth, which depends on the bath parameters.
Ref.~\cite{DAbbruzzo2022} provided numerical evidence that a similar logarithmic growth holds for the fermionic logarithmic negativity (FLN): this signals that the steady state possesses nontrivial entanglement content, at least within the framework of the Lindblad equation and for models that can be mapped to a free-fermion scenario.
Similar results are found in Ref.~\cite{Eisler2023}, which demonstrated logarithmic entanglement scaling in a steady-state ensemble similar to the one discussed in Ref.~\cite{DAbbruzzo2022}.

The master equation employed in Ref.~\cite{DAbbruzzo2022} takes the Lindblad form, with Lindblad operators that are constructed from the eigenstates of the system and thus are nonlocal in space~\cite{DAbbruzzo2021}.
Its microscopic derivation, besides the usual assumptions of weak coupling and Markovianity, relies on the so-called secular approximation: it requires the system-bath coupling to be small, compared to the level spacing of the system Hamiltonian spectrum~\cite{Breuer2002,Rivas2011}.
Secular master equations have been successfully applied over the decades in a variety of settings, most notably in quantum optics.
However, since in a quantum many-body system the level spacing typically decays exponentially with size, the validity of the secular approximation can become questionable.
As a consequence, a considerable amount of research has been recently devoted to build nonsecular master equations.
Arguably, the most prominent example of nonsecular master equation is the Redfield equation~\cite{Breuer2002,Redfield1965}, even though it has an important limitation: it does not necessarily lead to a completely positive dynamics~\cite{Dumcke1979}.
To overcome this issue, several directions have been explored, such as dynamical coarse graining~\cite{Schaller2008}, refined weak-coupling limit~\cite{Rivas2017}, partial secular approximation~\cite{Farina2019}, universal Lindblad equation (ULE)~\cite{Nathan2020,Davidovic2022}, and numerical regularization techniques~\cite{DAbbruzzo2023,DAbbruzzo2024} (see also Refs.~\cite{McCauley2020,Mozgunov2020,Potts2021,Becker2021,Trushechkin2021} for other approaches).

In this paper we investigate whether the steady-state entanglement scaling derived in Ref.~\cite{DAbbruzzo2022}, as measured by the FLN, survives beyond the framework of the nonlocal Lindblad equation.
Specifically, we focus on the predictions of the Redfield equation and the ULE.
Ref.~\cite{Prosen2010} hinted at a possible extensive character of the mutual information in the steady state predicted by the Redfield equation, but a more thorough analysis of the behavior of the entanglement is still missing.
We show that both approaches lead to logarithmic steady-state entanglement scaling, at least for intermediate-size subsystems.
Interestingly, upon increasing such size, the negativity crosses over to superlogarithmic scaling: this effect is shown to be more dramatic when increasing the system-bath coupling or the difference between the chemical potentials of the baths.

The paper is structured as follows.
We start with Sec.~\ref{sec:setting}, where we introduce a tight-binding chain coupled to thermal baths at its edges.
Sec.~\ref{sec:ME} is dedicated to the description of the Redfield equation (Sec.~\ref{sec:redfield}) and the ULE (Sec.~\ref{sec:ule});
we also discuss how to recover the nonlocal Lindblad equation employed in Refs.~\cite{DAbbruzzo2021,DAbbruzzo2022} (Sec.~\ref{sec:rotating}).
Then, in Sec.~\ref{sec:entanglement} we define the FLN as a measure of entanglement and discuss how to compute it in our setting.
In Sec.~\ref{sec:numerics} we proceed with a discussion of numerical results:
in Sec.~\ref{sec:steady-num} we first introduce the precise employed setup, whereas in Sec.~\ref{sec:neg-numerics} we show that in the weak-coupling regime the three master equations give the same scaling for the steady-state negativity;
in Sec.~\ref{sec:deviations} we address the superlogarithmic scaling effect away from the weak-coupling regime and in the thermodynamic limit.
Finally, in Sec.~\ref{sec:conclusions} we draw our conclusions.
The Appendix~\ref{app:majorana_lindblad} contains a derivation of the effective Lindblad form that can be obtained from the Redfield equation and the ULE, used in Sec.~\ref{sec:ME}.


\section{Tight-binding chain with boundary driving} \label{sec:setting}

\begin{figure}[t]
    \centering
    \includegraphics[width=\columnwidth]{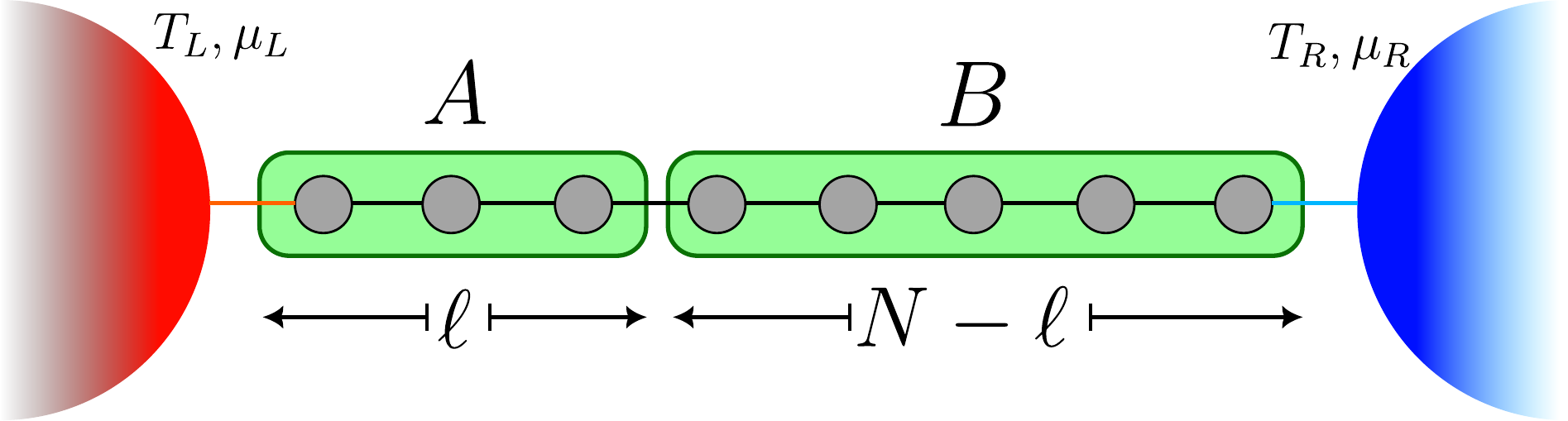}
    \caption{
        Setup used in this work (see Secs.~\ref{sec:setting} and~\ref{sec:entanglement}).
        A fermionic chain of $N$ sites is coupled at its edges to two fermionic thermal baths at temperatures $T_L$, $T_R$ and chemical potentials $\mu_L, \mu_R$.
        The chain is then partitioned into the two complementary regions $A$ and $B$ of sizes $\ell$ and $N-\ell$, respectively.
        We are interested in the scaling of the entanglement between $A$ and $B$, as measured by the FLN, in the thermodynamic limit $N,\ell \to \infty$.
    }
    \label{fig:setup}
\end{figure}

In this paper we consider a one-dimensional quantum system $\mathcal{S}$ in contact with two thermal reservoirs at its edges: such setup is schematically depicted in Fig.~\ref{fig:setup}.
The bulk of $\mathcal{S}$ is described by a tight-binding chain Hamiltonian with open boundary conditions:
\begin{equation} \label{eq:tight_binding}
    H_\mathcal{S} = - J \sum_{n=1}^N \qty[a_n^\dagger a_{n+1} + a_{n+1}^\dagger a_n + h \, a^\dagger_n a_n]
\end{equation}
with $J,h \in \mathbb{R}$ setting the energy scales of the system.
Here $a_n$ and $a^\dagger_n$ are fermionic annihilation and creation operators, satisfying
\begin{subequations}
    \begin{gather}
        \{ a_n, a_m^\dagger \} = \delta_{nm}, \\
        \{ a_n, a_m \} = \{ a_n^\dagger, a_m^\dagger \} = 0.
    \end{gather}
\end{subequations}
The boundary conditions are imposed by requiring $a_{N+1}^{(\dagger)} \equiv 0$.
Moreover, in the following we measure energetic quantities in units of $J$, hence we effectively fix $J=1$.
We also work in units of $\hbar = k_B = 1$.

The environment $\mathcal{E}$ consists of two free-fermionic thermal reservoirs described by the Hamiltonian $H_\mathcal{E} = H_\mathcal{E}^L + H_\mathcal{E}^R$, with
\begin{equation} \label{eq:H_E}
    H_\mathcal{E}^\alpha = \sum_{r=1}^\infty \epsilon_{\alpha,r} c_{\alpha,r}^\dagger c_{\alpha,r},
    \quad
    \alpha \in \{L,R\}.
\end{equation}
The indices $L$ and $R$ denote the ``left'' and ``right'' baths, respectively.
Here $\epsilon_{\alpha,r} \geq 0$ is the nonnegative energy of mode $r$ of bath $\alpha$, and $c_{\alpha,r},c_{\alpha,r}^\dagger$ are fermionic annihilation and creation operators satisfying
\begin{subequations}
    \begin{gather}
        \{ c_{\alpha,r}, c_{\beta,s}^\dagger \} = \delta_{\alpha\beta} \delta_{rs}, \\
        \{ c_{\alpha,r}, c_{\beta,s} \} = \{ c_{\alpha,r}^\dagger, c_{\beta,s}^\dagger \} = 0.
    \end{gather}
\end{subequations}
Both the reservoirs are prepared in a thermal state, each characterized by a temperature $T_\alpha$ and a chemical potential $\mu_\alpha$.
If we denote with $\rho_\mathcal{E}$ the density operator describing the environment, we have the following relations:
\begin{subequations} \label{eq:c_correlators}
    \begin{gather}
        \langle c_{\alpha,r} \rangle = \langle c_{\alpha,r}^\dagger \rangle = 0, \\
        \langle c_{\alpha,r}^\dagger c_{\beta,s} \rangle = \delta_{\alpha\beta} \, \delta_{rs} \, f_\alpha(\epsilon_{\alpha,r}), \\
        \langle c_{\alpha,r} c_{\beta,s}^\dagger \rangle = \delta_{\alpha\beta} \, \delta_{rs} \, [1-f_\alpha(\epsilon_{\alpha,r})],
    \end{gather}
\end{subequations}
where $\langle \cdot \rangle \coloneqq \Tr[ \, \cdot \, \rho_\mathcal{E}]$ and $f_\alpha(x)$ is the Fermi-Dirac distribution associated with bath $\alpha$, given by
\begin{equation} \label{eq:FD}
    f_\alpha(x) \coloneqq \frac{1}{1 + e^{(x - \mu_\alpha)/T_\alpha}} =
    \frac{1}{2} \qty[ 1 - \tanh(\frac{x - \mu_\alpha}{2T_\alpha}) ].
\end{equation}

Finally, for what concerns the system-bath interaction, we take the left and right reservoirs to interact only with the first and last site of the chain, respectively.
We take the Hamiltonian governing such interaction to be a bilinear function of the ladder operators of both the system and the environment:
\begin{equation} \label{eq:interaction}
    H_I = X_L \otimes Y_L + X_R \otimes Y_R,
\end{equation}
where
\begin{subequations}
    \begin{gather}
        X_L = a_1 + a_1^\dagger, \quad X_R = a_N + a_N^\dagger, \label{eq:X} \\
        Y_\alpha = \sum_{r=1}^\infty g_{\alpha,r} \qty( c_{\alpha,r} + c_{\alpha,r}^\dagger ),
        \quad \alpha \in \{L,R\}, \label{eq:Y}
    \end{gather}
\end{subequations}
and $g_{\alpha,r}$ is the interaction strength associated with mode $r$ of bath $\alpha$.
The bilinearity assumption will allow us to characterize the dynamics in terms of two-point correlation functions only (see Sec.~\ref{sec:ME}).
Other choices lead to a much more complex scenario that will not be considered here
[see, e.g., Ref.~\cite{Ng2015} for an example where an interaction similar to Eq.~\eqref{eq:interaction} is employed].
The total Hamiltonian is:
\begin{equation}
	H = H_0 + H_I \equiv H_\mathcal{S} \otimes \mathbbm{1} + \mathbbm{1} \otimes H_\mathcal{E}+H_I,
\end{equation}
with $H_0$ being the free term.

Before proceeding, we observe that the Hamiltonian $H_\mathcal{S}$ in Eq.~\eqref{eq:tight_binding} is straightforwardly diagonalized as~\cite{Lieb1961}
\begin{equation} \label{eq:E}
    H_\mathcal{S} = \sum_{k=1}^N E_k d_k^\dagger d_k + E_0,
\end{equation}
where the single-particle energy dispersion $E_k$ is
\begin{equation}
    E_k = -h - 2\cos( \frac{\pi k}{N+1} ),
\end{equation}
while $E_0$ is an irrelevant constant and $d_k$ is a fermionic annihilation operator defined by
\begin{equation} \label{eq:bogoliubov}
    a_n = \sum_{k=1}^N U_{n k} d_k,
    \quad
    U_{n k} = \sqrt{\frac{2}{N+1}} \sin(\frac{\pi n k}{N+1}).
\end{equation}
From these relations, one can show that $H_\mathcal{S}$ exhibits ground-state criticality in the thermodynamic limit ${N \to \infty}$ for $|h| \leq 2$, where it is described by a conformal field theory (CFT) with central charge $c = 1$~\cite{DiFrancesco1997,Its2005}.
In the following, we always assume $|h| \leq 2$.


\section{Redfield equation and ULE} \label{sec:ME}

Given the setting described in Sec.~\ref{sec:setting}, our main goal is to describe entanglement properties of the steady state of the chain.
First, one has to determine the evolution of the chain density matrix $\rho(t) = \Tr_\mathcal{E}[\rho_{\mathcal{SE}}(t)]$, where $\rho_{\mathcal{SE}}(t)$ is the state of the chain and the environment combined.
Clearly, $\rho$ satisfies the equation
\begin{equation} \label{eq:full-rho}
	\frac{d\rho(t)}{dt} = -i \Tr_\mathcal{E} [H, \rho_{\mathcal{SE}}(t)].
\end{equation}
As is well known, in general it is a challenging task to obtain a full solution to Eq.~\eqref{eq:full-rho}.
Here we rely on the assumption of weak system-environment coupling, which is a standard procedure in the theory of open quantum systems~\cite{Breuer2002,Rivas2011}.
The common way of performing such approximation leads to the Redfield equation~\cite{Redfield1965,Breuer2002,Rivas2011}, described in Sec.~\ref{sec:redfield}.
On the other hand, in Sec.~\ref{sec:ule} we describe the ULE~\cite{Nathan2020,Davidovic2022}, a recently derived completely-positive master equation that does not need the secular assumption, contrary to the Redfield equation.


\subsection{Redfield master equation} \label{sec:redfield}

At the lowest nontrivial order in the coupling between system and environment, from Eq.~\eqref{eq:full-rho} one finds the Redfield equation~\cite{Redfield1965,Breuer2002,Rivas2011}
\begin{equation} \label{eq:redfield}
    \frac{d\widetilde{\rho}(t)}{dt} = \sum_{\alpha,\beta} \int_0^\infty d\tau \, C_{\alpha\beta}(\tau) [\widetilde{X}_\beta(t-\tau) \widetilde{\rho}(t), \widetilde{X}_\alpha(t)] + \text{H.c.},
\end{equation}
where $\alpha,\beta \in \{L,R\}$ are bath indices, the tilde denotes operators in interaction picture, meaning that ${ \widetilde{X}(t) \coloneqq e^{i H_0 t} X(t) e^{-i H_0 t} }$, and
\begin{equation}
    C_{\alpha\beta}(\tau) \coloneqq \langle \widetilde{Y}_\alpha(\tau) Y_\beta \rangle
\end{equation}
is the environment correlation function.
Making use of Eqs.~\eqref{eq:c_correlators} and~\eqref{eq:Y}, one obtains ${C_{\alpha\beta}(\tau) = \delta_{\alpha\beta} C_{\alpha}(\tau)}$, with
\begin{equation} \label{eq:c}
	C_{\alpha}(\tau) = \int_{-\infty}^\infty \!\! d x \, \mathcal{J}_\alpha(x) \qty{ e^{ix\tau} f_\alpha(x) + e^{-ix\tau} [1-f_\alpha(x)] },
\end{equation}
where we introduced the spectral density function
\begin{equation} \label{eq:J}
	\mathcal{J}_\alpha(x) \coloneqq \sum_{r=1}^\infty g_{\alpha,r}^2 \delta(x-\epsilon_{\alpha,r}).
\end{equation}
Note that, since $\epsilon_{\alpha,r} \geq 0$, we must have $\mathcal{J}_\alpha(x) = 0$
for $x < 0$.

In Appendix~\ref{app:majorana_lindblad} we show that Eq.~\eqref{eq:redfield} can be written in a pseudo-Lindblad form in terms of the normal-mode Majorana operators, defined as
\begin{equation} \label{eq:normal_majorana}
    m_{2k-1} \coloneqq \frac{1}{\sqrt{2}} \qty(d_k^\dagger + d_k),
    \quad
    m_{2k} \coloneqq \frac{i}{\sqrt{2}} \qty(d_k^\dagger - d_k),
\end{equation}
with $d_k$ being the fermionic modes introduced in Eq.~\eqref{eq:E}.
Specifically, one obtains
\begin{align} \label{eq:lindblad}
    \frac{d\rho(t)}{dt} =& - \sum_{c,d=1}^{2N} L_{cd} \qty[m_d m_c, \rho(t)]  \nonumber \\
    &+ \sum_{c,d=1}^{2N} M_{cd} \qty[ 2m_c \rho(t) m_d - \qty{m_d m_c, \rho(t)} ].
\end{align}
Note that the operators in Eq.~\eqref{eq:lindblad} are not in interaction picture anymore.
Here $L$ and $M$ are $2N \times 2N$ complex matrices such that their entries at positions ${(2k-1, 2q-1)}$, ${(2k-1,2q)}$, ${(2k,2q-1)}$, and ${(2k,2q)}$, for ${k,q \in \{1,\ldots,N\}}$, are arranged in the $2 \times 2$ blocks $\mathcal{L}_{kq}$ and $\mathcal{M}_{kq}$ given by
\begin{subequations} \label{eq:LM}
    \begin{gather}
        \mathcal{L}_{k q} = \frac{1}{2} \begin{pmatrix}
            \mathcal{A}^r_{k q} - [\mathcal{A}^r_{q k}]^* & \delta_{kq} E_k - [\mathcal{D}^r_{q k}]^* \\
            -\delta_{kq} E_k + \mathcal{D}^r_{k q} & 0
        \end{pmatrix}, \\
        \mathcal{M}_{k q} = \frac{1}{2} \begin{pmatrix}
            \mathcal{A}^r_{k q} + [\mathcal{A}^r_{q k}]^* & [\mathcal{D}^r_{q k}]^* \\
            \mathcal{D}^r_{k q} & 0
        \end{pmatrix}.
    \end{gather}
\end{subequations}
We introduced the quantities $\mathcal{A}^r_{kq}$ and $\mathcal{D}^r_{kq}$ as
\begin{subequations} \label{eq:red_AD}
    \begin{gather}
        \mathcal{A}^r_{k q} = \sum_\alpha \varphi_{\alpha,k} \varphi_{\alpha,q} \qty[ \gamma_\alpha(E_k) + \gamma_\alpha(-E_k) ], \\
        \mathcal{D}^r_{k q} = i \sum_\alpha \varphi_{\alpha,k} \varphi_{\alpha,q} \qty[ \gamma_\alpha(E_k) - \gamma_\alpha(-E_k) ],
    \end{gather}
\end{subequations}
where
\begin{subequations}
    \begin{gather}
	    \varphi_{L,k} \coloneqq U_{1,k},
        \quad
	\varphi_{R,k} \coloneqq U_{N,k}, \label{eq:phi} \\
	\gamma_\alpha(\omega) \coloneqq \int_0^\infty d\tau \, C_{\alpha}(\tau) e^{i\omega\tau}, \label{eq:gamma}
    \end{gather}
\end{subequations}
with $U_{1,k}$ and $U_{N,k}$ being defined in Eq.~\eqref{eq:bogoliubov}.
Note that, in general, $\mathcal{A}^r_{kq}$ and $\mathcal{D}^r_{kq}$ are not diagonal matrices.

Using the expression for $C_{\alpha}(\tau)$ in Eq.~\eqref{eq:c} and the well-known formula
\begin{equation} \label{eq:sokplem}
    \int_0^\infty d\tau \, e^{ix\tau} = \pi \delta(x) + i \mathbb{P} \frac{1}{x},
\end{equation}
with $\mathbb{P}$ being the Cauchy principal value sign, we can rewrite Eq.~\eqref{eq:gamma} as
\begin{equation}
	\gamma_\alpha(\omega) = \pi \hat{C}_{\alpha}(\omega) +
	i \mathbb{P} \! \int_{-\infty}^\infty d x \, \frac{\hat{C}_{\alpha}(x)}{\omega - x},
\end{equation}
where $\hat{C}_{\alpha}(\omega) \coloneqq \int_{-\infty}^\infty (d\tau/2\pi) C_{\alpha}(\tau) e^{i\omega\tau}$ is the Fourier transform of $C_{\alpha}(\tau)$, given by
\begin{equation} \label{eq:calpha}
	\hat{C}_{\alpha}(\omega) = \mathcal{J}_\alpha(-\omega) f_\alpha(-\omega) + \mathcal{J}_\alpha(\omega) [1-f_\alpha(\omega)].
\end{equation}

Instead of working with the density matrix $\rho$, it is convenient to recast Eq.~\eqref{eq:lindblad} into a differential equation for the correlation matrix
\begin{equation}
	G_{ab} \coloneqq -i \Tr( [m_a, m_b] \rho).
\end{equation}
It is straightforward to use the adjoint of Eq.~\eqref{eq:lindblad} to obtain the following differential continuous Lyapunov equation~\cite{Gajic1995,Purkayastha2022,Barthel2022}:
\begin{equation} \label{eq:lyapunov}
    \frac{dG}{dt} + PG + GP^T = Q,
\end{equation}
where
\begin{equation} \label{eq:PQ}
	P_{cd} = 2 \Re[M_{cd}-L_{cd}],
    \quad
    Q_{cd} = -4 \Im[M_{cd}].
\end{equation}

Equation~\eqref{eq:lyapunov} describes the dynamics of the Majorana operators $m_{2k-1},m_{2k}$ defined from the Fourier-transformed fermions $d_k$ [cf. Eq.~\eqref{eq:normal_majorana}].
However, to determine entanglement-related quantities, one has to compute the dynamics of the Majorana correlation matrix in real space, which is defined as
\begin{equation} \label{eq:gamma-space}
    \Gamma_{ab}(\rho) \coloneqq -i \Tr([w_a,w_b]\rho),
\end{equation}
where $w_a$ are the Majorana operators
\begin{equation}
	w_{2n-1} \coloneqq \frac{1}{\sqrt{2}} \qty( a_n + a^\dagger_n ),
    \quad
	w_{2n} \coloneqq \frac{i}{\sqrt{2}} \qty( a_n^\dagger - a_n )
\end{equation}
defined from the original fermions $a_n,a^\dagger_n$ in Eq.~\eqref{eq:tight_binding}.
The change of basis between the two Majorana operators $m_a$ and $w_a$ is given in terms of the matrix $U_{nk}$ in Eq.~\eqref{eq:bogoliubov}.
Specifically, one finds
\begin{equation} \label{eq:change}
	\begin{pmatrix}
        w_{2n-1} \\ w_{2n}
    \end{pmatrix}
    = \sum_{k=1}^N
    \begin{pmatrix}
        U_{n k} & 0 \\
        0 & U_{n k}
    \end{pmatrix}
    \begin{pmatrix}
        m_{2k-1} \\ m_{2k}
    \end{pmatrix},
\end{equation}
which is equivalently written as
\begin{equation}
	w_a = \sum_{b=1}^{2N} K_{ab} m_b,
    \quad a \in \{1,\ldots,2N\},
\end{equation}
where $K$ is defined as
\begin{equation}
    K = U \otimes \mathbbm{1}_2,
\end{equation}
and $\mathbbm{1}_2$ is the $2 \times 2$ identity matrix.
Clearly, $K$ is an orthogonal matrix.
Thus, from Eq.~\eqref{eq:lyapunov}, using the fact that $\Gamma=K G K^T$, we obtain the dynamics of the real-space Majorana correlation matrix $\Gamma$ as
\begin{equation} \label{eq:lyapunov_Gamma}
    \frac{d\Gamma}{dt} + \widetilde{P} \Gamma + \Gamma \widetilde{P}^T = \widetilde{Q},
\end{equation}
where
\begin{equation}
    \widetilde{P} = K P K^T,
    \quad
    \widetilde{Q} = K Q K^T,
\end{equation}
and $P,Q$ are defined in Eq.~\eqref{eq:PQ}.
The existence of Eq.~\eqref{eq:lyapunov_Gamma} reflects that the property of a state of being described by a two-point correlation matrix only, i.e., Gaussianity, is preserved by the dynamics.


\subsection{Recovering the nonlocal Lindblad equation} \label{sec:rotating}

Before proceeding, we show how to recover the nonlocal Lindblad equation derived in Ref.~\cite{DAbbruzzo2021} and employed in Ref.~\cite{DAbbruzzo2022}.
The nonlocal Lindblad equation relies on a secular approximation~\cite{Breuer2002,Rivas2011}, which in the present formalism consists in neglecting all the contributions with $k \neq q$ in Eqs.~\eqref{eq:LM}.
As a consequence, the matrices $\mathcal{A}^r_{kq}$ and $\mathcal{D}^r_{kq}$ in Eqs.~\eqref{eq:red_AD} become diagonal and the coefficient matrices $P$ and $Q$ in Eq.~\eqref{eq:PQ} become block-diagonal:
\begin{subequations} \label{eq:PQ_nonlocal}
    \begin{align}
	    \label{eq:P-nonlocal}
        P &= \bigoplus_{k=1}^N
        \begin{pmatrix}
            2 \Re \mathcal{A}^r_{k k} & -E_k + 2 \Re \mathcal{D}^r_{k k} \\
            E_k & 0
        \end{pmatrix}, \\
	    \label{eq:Q-nonlocal}
        Q &= \bigoplus_{k=1}^N
        \begin{pmatrix}
            0 & 2 \Im \mathcal{D}^r_{k k} \\
            - 2 \Im \mathcal{D}^r_{k k} & 0
        \end{pmatrix}.
    \end{align}
\end{subequations}
The Lyapunov equation~\eqref{eq:lyapunov} can then be solved explicitly.
In the steady state, where $dG/dt = 0$, one can use the ansatz
\begin{equation} \label{eq:ck}
    G = \bigoplus_{k=1}^N
    \begin{pmatrix}
        0 & z_k \\ -z_k & 0
    \end{pmatrix},
\end{equation}
which leads to
\begin{equation}
	z_k = \frac{\Im \mathcal{D}^r_{k k}}{\Re \mathcal{A}^r_{k k}} = \frac{\sum_\alpha
	\varphi_{\alpha,k}^2 [\hat{C}_{\alpha}(E_k) - \hat{C}_{\alpha}(-E_k)]}
	{\sum_\alpha \varphi_{\alpha,k}^2 [\hat{C}_{\alpha}(E_k) + \hat{C}_{\alpha}(-E_k)]},
\end{equation}
One can check that this is the same result discussed in Refs.~\cite{DAbbruzzo2021,DAbbruzzo2022}.


\subsection{Universal Lindblad Equation} \label{sec:ule}

The so-called ULE~\cite{Nathan2020,Davidovic2022} in the interaction picture takes the following form
\begin{multline} \label{eq:ULE}
    \frac{d\widetilde{\rho}(t)}{dt} = \sum_\alpha \int_{-\infty}^{\infty} d s \int_{-\infty}^\infty d s' \, \theta(s-s') \\
    \times g_\alpha(s-t) g_\alpha(t-s') [\widetilde{X}_\alpha(s') \widetilde{\rho}(t), \widetilde{X}_\alpha(s)] + \text{H.c.},
\end{multline}
where $g_\alpha$ is the so-called ``jump'' correlation function, defined as
\begin{equation} \label{eq:jump}
	g_\alpha(\tau) \coloneqq \frac{1}{\sqrt{2\pi}} \int_{-\infty}^\infty d\omega \sqrt{\hat{C}_{\alpha}
	(\omega)} e^{-i\omega\tau},
\end{equation}
In Eq.~\eqref{eq:ULE} we already assumed that $C_{\alpha\beta}(\tau) = \delta_{\alpha\beta} C_{\alpha}(\tau)$.

We can write Eq.~\eqref{eq:ULE} in Lindblad form as we did for the Redfield equation in Sec.~\ref{sec:redfield}.
The result, which we derive in Appendix~\ref{app:majorana_lindblad}, is the same as Eq.~\eqref{eq:lindblad}, where now
\begin{subequations} \label{eq:LM_ULE}
    \begin{gather}
        \mathcal{L}_{k q} =
        \begin{pmatrix}
            i \Im \mathcal{A}^u_{k q} & \delta_{kq} E_k/2 + \Re \mathcal{C}^u_{k q} \\
            -\delta_{kq} E_k/2 - \Re \mathcal{C}^u_{q k} & i \Im \mathcal{B}^u_{k q}
        \end{pmatrix}, \\
        \mathcal{M}_{k q} =
        \begin{pmatrix}
            \Re \mathcal{A}^u_{k q} & i \Im \mathcal{C}^u_{k q} \\
            -i \Im \mathcal{C}^u_{q k} & \Re \mathcal{B}^u_{k q}
        \end{pmatrix},
    \end{gather}
\end{subequations}
and $\mathcal{A}^u_{kq}$, $\mathcal{B}^u_{kq}$, $\mathcal{C}^u_{kq}$ are defined as $(\sigma, \sigma' \in \{+1,-1\}$)
\begin{subequations} \label{eq:ULE_ABC}
    \begin{align}
        \mathcal{A}^u_{k q} &= \frac{1}{2} \sum_{\sigma,\sigma'} \sum_\alpha \varphi_{\alpha,k} \varphi_{\alpha,q} \, \mathcal{I}_\alpha(\sigma E_k, \sigma' E_q), \\
        \mathcal{B}^u_{k q} &= \frac{1}{2} \sum_{\sigma,\sigma'} \sum_\alpha \varphi_{\alpha,k} \varphi_{\alpha,q} \, (-\sigma \sigma') \mathcal{I}_\alpha(\sigma E_k, \sigma' E_q), \\
	    \mathcal{C}^u_{k q} &= \frac{i}{2} \sum_{\sigma,\sigma'} \sum_\alpha \varphi_{\alpha,k} \varphi_{\alpha,q} \, (-\sigma') \mathcal{I}_\alpha(\sigma E_k, \sigma' E_q).
    \end{align}
\end{subequations}
We also introduced $\mathcal{I}_\alpha(\omega,\omega')$ as
\begin{multline} \label{eq:I_better}
	\mathcal{I}_\alpha(\omega, \omega') = \pi \sqrt{ \hat{C}_{\alpha}(-\omega) \hat{C}_{\alpha}(\omega') } \\
	- i \mathbb{P} \! \int_{-\infty}^\infty \frac{d x}{x} \sqrt{ \hat{C}_{\alpha}(x-\omega) \hat{C}_{\alpha}(x+\omega') }.
\end{multline}
Now, the same Lyapunov equation~\eqref{eq:lyapunov} holds true.
The matrices $P$ and $Q$ are the same as in Eq.~\eqref{eq:PQ} with $L,M$ obtained from Eqs.~\eqref{eq:LM_ULE}.
Similarly to the Redfield equation, $P$ and $Q$ are not block-diagonal matrices, in general.


\section{Fermionic logarithmic negativity} \label{sec:entanglement}

Let us now consider the bipartition of the chain in two complementary connected regions $A$ and $B$ (see Fig.~\ref{fig:setup}).
The two regions correspond to the sites $\{1,\ldots,\ell\}$ and $\{\ell + 1, \ldots, N\}$, respectively.
We are interested in quantifying the entanglement between $A$ and $B$, given the density operator $\rho$ of the whole chain.

Crucially, due to the presence of the thermal reservoirs, the full-chain density matrix $\rho$ is mixed.
This implies that neither the von Neumann entropy nor the R\'{e}nyi entropies are proper entanglement measures~\cite{Groisman2005}.
A popular alternative, which is instead a computable measure of entanglement, is the logarithmic negativity~\cite{Vidal2002}:
\begin{equation}
    \mathscr{E}(\rho, A) \coloneqq \ln \Tr |\rho^{T_A}|,
\end{equation}
where $|X| \coloneqq \sqrt{X^\dagger X}$, and $\rho^{T_A}$ is the partial transpose of $\rho$ with respect to subsystem $A$, defined by
\begin{equation}
    \mel*{ e^{(A)}_i e^{(B)}_j }{\rho^{T_A}}{ e^{(A)}_k
    e^{(B)}_q } = \mel*{ e^{(A)}_k e^{(B)}_j }{\rho}{ e^{(A)}_i e^{(B)}_q },
\end{equation}
for any orthonormal basis $\{\ket*{e^{\scriptscriptstyle(A)}_i}\}$ of $A$ and $\{\ket*{e^{\scriptscriptstyle(B)}_i}\}$ of $B$.
For free-boson systems, if $\rho$ is a Gaussian density matrix, $\rho^{T_A}$ remains Gaussian.
This means that $\mathscr{E}$ can be computed from the covariance matrix of $\rho$~\cite{Audenaert2002}.
In contrast, for free-fermion systems, this is not true anymore.

In fact, let us rewrite the Majorana correlation matrix $\Gamma(\rho)$ [cf. Eq.~\eqref{eq:gamma-space}] as
\begin{equation}
    \Gamma(\rho) \equiv \begin{pmatrix}
        \Gamma^{AA}(\rho) & \Gamma^{AB}(\rho) \\
        \Gamma^{BA}(\rho) & \Gamma^{BB}(\rho)
    \end{pmatrix},
\end{equation}
where $\Gamma^{XY}$ is the Majorana correlation matrix where the row and column indices are restricted to subsystems $X$ and $Y$, respectively.
One can show that~\cite{Eisler2015}
\begin{equation}
	\label{eq:decomp}
    \rho^{T_A} = \frac{1-i}{2} O_+ + \frac{1+i}{2} O_-,
\end{equation}
where $O_\pm$ are Gaussian operators with corresponding Majorana correlation matrices
\begin{equation} \label{eq:G_pm}
    \Gamma(O_\pm) \equiv \Gamma_\pm = \begin{pmatrix}
        -\Gamma^{AA}(\rho) & \pm i \Gamma^{AB}(\rho) \\
        \pm i \Gamma^{BA}(\rho) & \Gamma^{BB}(\rho)
    \end{pmatrix}.
\end{equation}
Unfortunately, as it is clear from Eq.~\eqref{eq:decomp}, $\rho^{T_A}$ is not Gaussian, which implies that its spectrum and the negativity cannot be computed efficiently.
However, from $O_\pm$ one can define the FLN ${\mathscr{E}}_F$ as
\begin{equation}
	\mathscr{E}_F(\rho, A) \coloneqq \ln \Tr \sqrt{O_+ O_-}.
\end{equation}
The FLN is an upper bound for the standard logarithmic negativity that is still a proper entanglement measure, and can be defined in terms of a partial time-reversal transformation of the density matrix~\cite{Shapourian2017,Eisert2018,Shapourian2019}.
Since $O_+O_-$ is a Gaussian operator, its spectrum and ${\mathscr E}_F$ can be computed from $\Gamma_\pm$.
One obtains~\cite{Eisert2018}
\begin{equation} \label{eq:negativity_renyi}
    \mathscr{E}_F(\rho, A) = \frac{1}{2} \qty[ S_{1/2}(\rho_\times) - S_2(\rho) ],
\end{equation}
where $S_\alpha(\rho)$ is the $\alpha$-R\'{e}nyi entropy
\begin{equation}
	S_\alpha(\rho) \coloneqq \frac{1}{1-\alpha} \ln \Tr[\rho^\alpha],
\end{equation}
and $\rho_\times = O_+ O_- / \Tr[O_+ O_-]$, characterized by the Majorana correlation matrix
\begin{equation} \label{eq:cross_correlation}
    \Gamma(\rho_\times) = -i\mathbbm{1} + i(\mathbbm{1}-i\Gamma_-)(\mathbbm{1}-\Gamma_+ \Gamma_-)^{-1} (\mathbbm{1}-i\Gamma_+).
\end{equation}
For a Gaussian state $\rho$, one can show that~\cite{Camilo2019}
\begin{equation} \label{eq:renyi_spectral}
    S_\alpha(\rho) = \frac{1}{1-\alpha} \sum_j \log\qty[ \qty( \frac{1-\nu_j}{2} )^\alpha + \qty( \frac{1+\nu_j}{2} )^\alpha ],
\end{equation}
where $\pm i \nu_j$ ($\nu_j \in \mathbb{R}$) are the eigenvalues of $\Gamma(\rho)$, being of this form because $\Gamma(\rho)$ is a real skew-symmetric matrix.


\section{Steady-state entanglement: Numerical results} \label{sec:numerics}

In this section we numerically investigate the scaling of the FLN in the steady state of the chain.
Specifically, in Sec.~\ref{sec:steady-num} we start with a brief discussion on how this quantity is calculated.
Then, in Sec.~\ref{sec:neg-numerics} we discuss numerical results for $\mathscr{E}_F$ obtained by employing the three different master equations introduced in Sec.~\ref{sec:ME} in the weak-coupling regime, showing that logarithmic scaling holds for values of $\ell$ that are not too large.
Finally, in Sec.~\ref{sec:deviations} we show that away from the weak-coupling regime, or in the limit $\ell \to \infty$, the Redfield and ULE approaches give superlogarithmic scaling, in contrast with the nonlocal Lindblad equation which always yields logarithmic increase.


\subsection{Obtaining the steady-state FLN} \label{sec:steady-num}

The first step of the procedure is the construction of the matrices $P$ and $Q$ in Eq.~\eqref{eq:PQ} [see also Eqs.~\eqref{eq:LM} for the Redfield equation, Eqs.~\eqref{eq:PQ_nonlocal} for the nonlocal Lindblad equation, and Eqs.~\eqref{eq:LM_ULE} for the ULE].
To proceed we should fix the spectral density~\eqref{eq:J} and the parameters $\mu_\alpha, T_\alpha$ of the reservoirs in~\eqref{eq:FD}.
Here we choose for both baths an Ohmic spectral density with Lorentz-Drude cutoff:
\begin{equation} \label{eq:J_lorentz}
	\mathcal{J}_\alpha(x) = \theta(x) \frac{\gamma x}{x^2 + \Omega^2},
    \quad
    \gamma, \Omega > 0,
\end{equation}
where the Heaviside theta function $\theta(x)$ reflects that $\epsilon_{\alpha,r} \geq 0$, with $\epsilon_{\alpha,r}$ being the single-particle energies of the reservoirs.
To guarantee the validity of the Born-Markov approximation~\cite{Breuer2002,Rivas2011}, one typically assumes that $\gamma \ll \Omega$~\cite{DAbbruzzo2023}, which corresponds to a weak-coupling regime.

The next step is to numerically solve the continuous Lyapunov equation~\eqref{eq:lyapunov_Gamma} for the Majorana correlation matrix $\Gamma$.
Here we are interested in the steady state arising at $t \to \infty$, where $d\Gamma/dt = 0$ and we are left with the matrix equation
\begin{equation} \label{eq:steady_state}
	\widetilde{P} \Gamma + \Gamma \widetilde{P}^T = \widetilde{Q},
\end{equation}
which can be solved with polynomially-scaling complexity $\mathcal{O}(N^3)$, for instance, by using the Bartels-Stewart algorithm~\cite{Bartels1972}.
Notice that, due to the Gaussianity of the state, the correlation matrix $\Gamma$ encodes complete information about the steady state.
This is not the case for interacting fermions, for which one would have to solve the master equation for $\rho$, with a computational cost that scales exponentially with $N$.

Importantly, one can prove that Eq.~\eqref{eq:steady_state} has a unique solution (i.e., the dynamics has a unique steady state) if and only if $\widetilde{P}$ and $-\widetilde{P}^T$ do not share any eigenvalue, meaning that $\lambda_i + \lambda_j \neq 0$ for any pair of eigenvalues $\lambda_i, \lambda_j$ of $\widetilde{P}$.
We checked that this condition is verified for our system.

Once we have $\Gamma$, we obtain the FLN as described in Sec.~\ref{sec:entanglement}.
Specifically, we consider the bipartition in Fig.~\ref{fig:setup}, in which region $A$ and $B$ correspond to the first $\ell$ sites and the last $N-\ell$ sites of the chain, respectively.
We first determine $\Gamma_\pm$ using Eq.~\eqref{eq:G_pm}, then we calculate $\Gamma(\rho_\times)$ using Eq.~\eqref{eq:cross_correlation}, and finally we obtain $\mathscr{E}_F$ by spectral decomposition using Eqs.~\eqref{eq:negativity_renyi} and~\eqref{eq:renyi_spectral}.


\subsection{Weak-coupling regime} \label{sec:neg-numerics}

\begin{figure}
    \centering
    \includegraphics[width=\columnwidth]{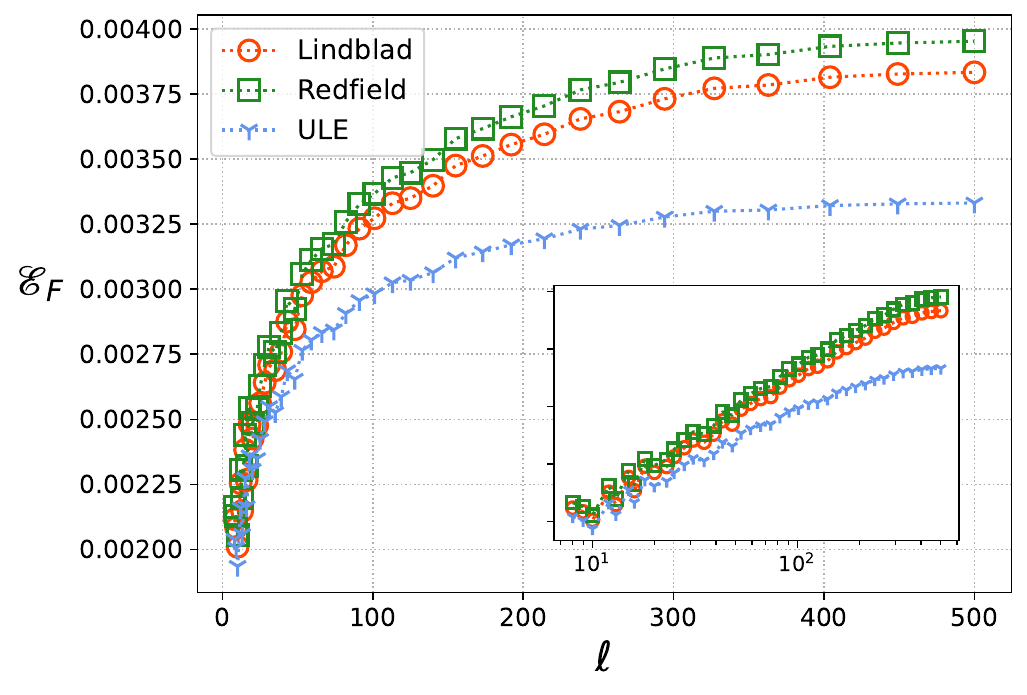}
    \caption{
        Steady-state FLN as a function of the size $\ell$ of region $A$ (see Fig.~\ref{fig:setup}), for $\ell \leq N/2$.
        The data are for the tight-binding chain with $N=1000$ and $h=1.1$, connected to two reservoirs with temperatures $T_L=10$, $T_R=15$ and chemical potentials $\mu_L=1$, $\mu_R=1.5$.
        Here we choose $\gamma=0.2$ and $\Omega=10$ for the spectral density [cf.~\eqref{eq:J_lorentz}].
        The inset shows the same data in logarithmic scale on the $x$-axis.
    }
    \label{fig:example}
\end{figure}

In Fig.~\ref{fig:example} we start by plotting the steady-state FLN $\mathscr{E}_F$ as a function of the size $\ell$ of region $A$.
We show results for $N=1000$ sites and $h=1.1$.
We also fix the bath parameters to $T_L=10$, $T_R=15$, $\mu_L=1$, and $\mu_R=1.5$.
For the spectral density in~\eqref{eq:J_lorentz} we choose $\Omega=10$ and $\gamma=0.2$, so that we are in the weak-coupling regime.
The various symbols in the figure correspond to results associated with different master equations.
Provided that $\ell$ is sufficiently smaller than $N$, the FLN exhibits logarithmic scaling as a function of $\ell$ for all the three master equations.
Even though the three approaches give different numerical results, the qualitative behavior of the FLN is the same.
For the nonlocal Lindblad equation, this is compatible with what was established in Ref.~\cite{DAbbruzzo2022}, but this was not obvious for the Redfield and the ULE approaches.

Inspired by the results of Ref.~\cite{DAbbruzzo2022}, we now consider how $\mathscr{E}_F$ depends on the so-called chord length $X_\ell$, which is defined as
\begin{equation} \label{eq:chord}
    {X}_\ell \coloneqq \frac{N}{\pi} \sin(\frac{\pi\ell}{N}),
    \quad
    \ell=1,\ldots,N.
\end{equation}
In systems described by a CFT, the scaling of the entanglement entropy in the limit $N,\ell \to \infty$ depends on $X_\ell$, and not on $\ell$ and $N$ separately~\cite{DiFrancesco1997,Calabrese2009a}.
Such a feature was also verified in ground states of spin chains described by the random singlet phase~\cite{Fagotti2011}.
Moreover, one can show that the FLN between two complementary intervals in a globally pure state becomes the R\'{e}nyi entropy of one of the intervals, which in CFTs exhibits a logarithmic scaling with the chord length~\cite{Shapourian2017}.
Furthermore, the mutual information between $A$ and $B$ exhibits scaling as a function of $X_\ell$~\cite{DAbbruzzo2022}.

\begin{figure}
    \centering
    \includegraphics[width=\columnwidth]{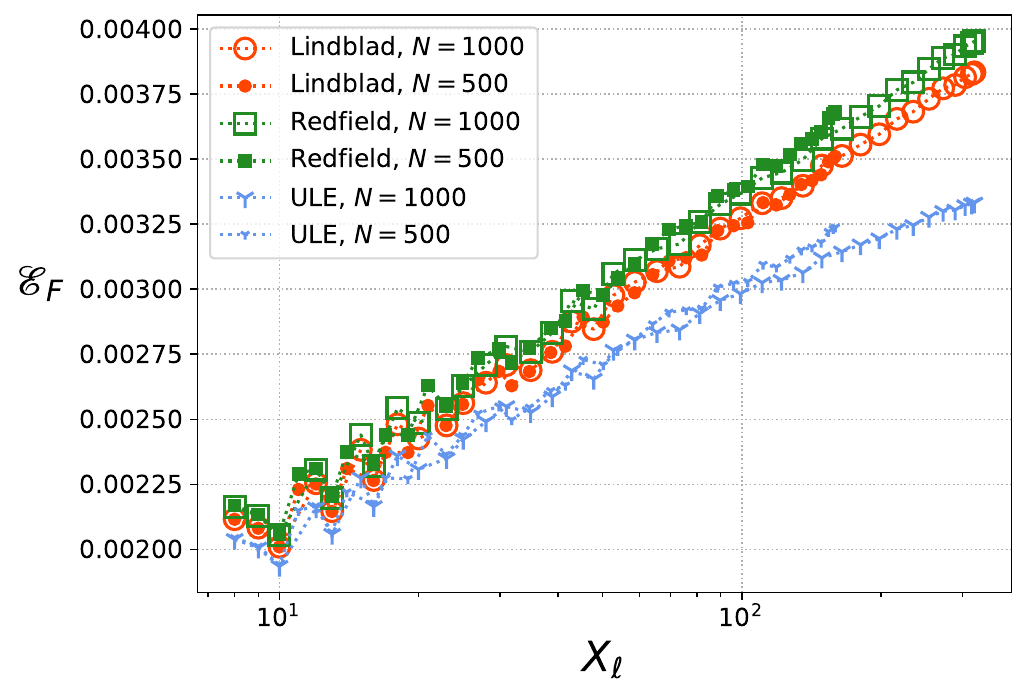}
    \caption{
        Steady-state FLN as a function of the chord length $X_\ell$ [cf.~\eqref{eq:chord}].
        Notice the logarithmic scale on the $x$-axis.
        The full symbols are the results for $N=500$, while the empty ones are for $N=1000$.
        The other parameters are the same as in Fig.~\ref{fig:example}.
    }
    \label{fig:example-1}
\end{figure}

In Fig.~\ref{fig:example-1} we plot the steady-state FLN as a function of $X_\ell$, using the same parameters of Fig.~\ref{fig:example}.
Numerical data exhibit logarithmic scaling for all the three master equations, even though the prefactor of such a logarithmic growth is different in the three cases.
This allows us to conclude that the logarithmic entanglement scaling observed in Ref.~\cite{DAbbruzzo2022} for the nonlocal Lindblad equation holds true also in the Redfield and ULE approaches, at least for weak coupling and moderately large subsystem size $\ell$.

However, there is another fact that can be discussed here.
If it is true that $\mathscr{E}_F$ scales as a function of $X_\ell$ only, the data at different values of $N$ should collapse on the same curve, at least in the thermodynamic limit $N, \ell \to \infty$.
In Fig.~\ref{fig:example-1} we also plot the results obtained by choosing another value for the chain length, $N=500$, in full symbols.
While we observe perfect scaling for the nonlocal Lindblad equation, the scenario is different for the Redfield and ULE approaches.
For instance, for the ULE the data for $N=500$ and $N=1000$ suggest collapse on the same curve up to $X_\ell \lesssim 10^2$.
Similar behavior occurs for the Redfield equation, even though the violations are significantly weaker.
While such violations of scaling collapse could be attributed to finite-size effects, we will show in Sec.~\ref{sec:deviations} that they rather signal a crossover to superlogarithmic scaling at asymptotically large $X_\ell$.

\begin{figure}
    \centering
    \includegraphics[width=\columnwidth]{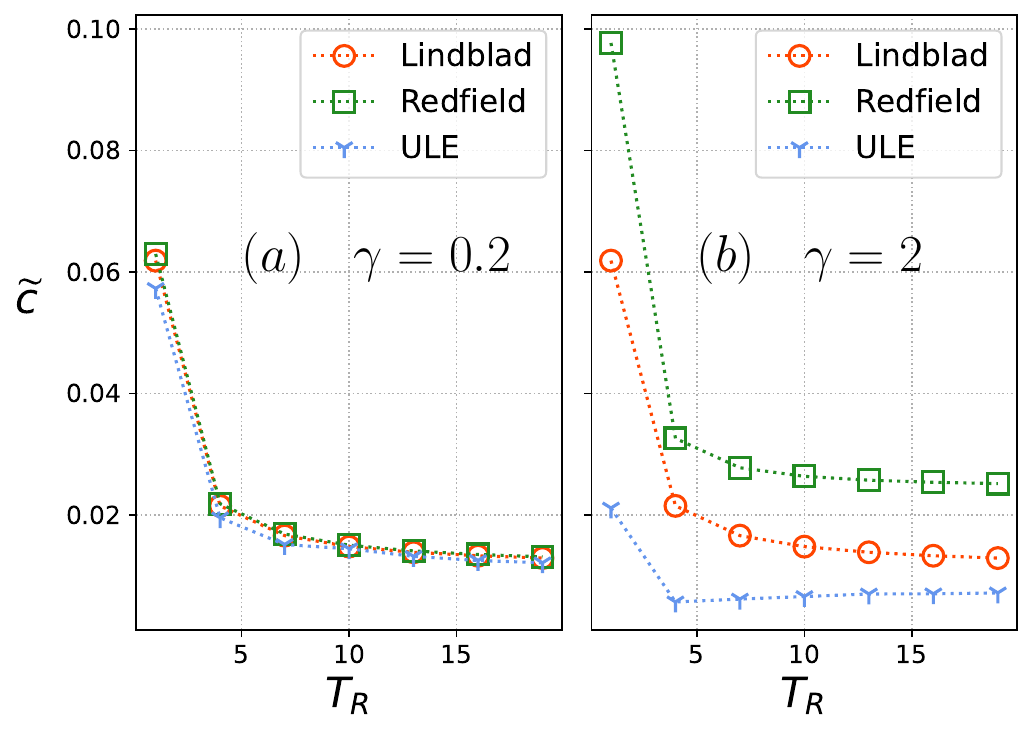}
    \caption{
	    Prefactor $\widetilde{c}$ of the logarithmic scaling of the steady-state FLN [cf. Eq.~\eqref{eq:e-X}] as a function of $T_R$ for fixed $T_L = 1$, calculated at $N=1000$.
        Panels (a) and (b) are for two values of the system-bath interaction strength $\gamma=0.2$ and $\gamma=2$, respectively, while the other parameters are fixed as in Fig.~\ref{fig:example}.
    }
    \label{fig:temperature}
\end{figure}

Before proceeding, we investigate the behavior of the prefactor of the putative logarithmic scaling at moderate values of $X_\ell$, obtained by fitting the data in Fig.~\ref{fig:example-1} with
\begin{equation} \label{eq:e-X}
    \mathscr{E}_F = \widetilde{c} \ln(X_\ell) + b,
\end{equation}
where $\widetilde{c}$ and $b$ are the fitting parameters.
In Fig.~\ref{fig:temperature} we plot $\widetilde{c}$ as a function of $T_R$, for fixed $T_L = 1$.
The two panels correspond to two different values of the coupling $\gamma$.
For $\gamma = 0.2$ [panel (a)], $\widetilde{c}$ is similar for the three master equations: this is expected, since by construction the three approaches become equivalent, in the limit $\gamma \to 0$.
At $T_R \approx T_L$, $\widetilde{c}$ is larger, and it exhibits a decreasing trend upon increasing $T_R$, which is qualitatively similar to the behavior of the mutual information derived in Ref.~\cite{DAbbruzzo2022}.
Note that, in principle, this could be confirmed analytically for the nonlocal Lindblad equation using the results of Refs.~\cite{DAbbruzzo2022, Eisler2023}.
On the other hand, for $\gamma = 2$ [panel (b)] the results obtained for the three master equations become significantly different from each other.


\subsection{Violations of logarithmic scaling in the Redfield and the ULE approaches} \label{sec:deviations}

\begin{figure*}
    \centering
    \includegraphics[width=.9\linewidth]{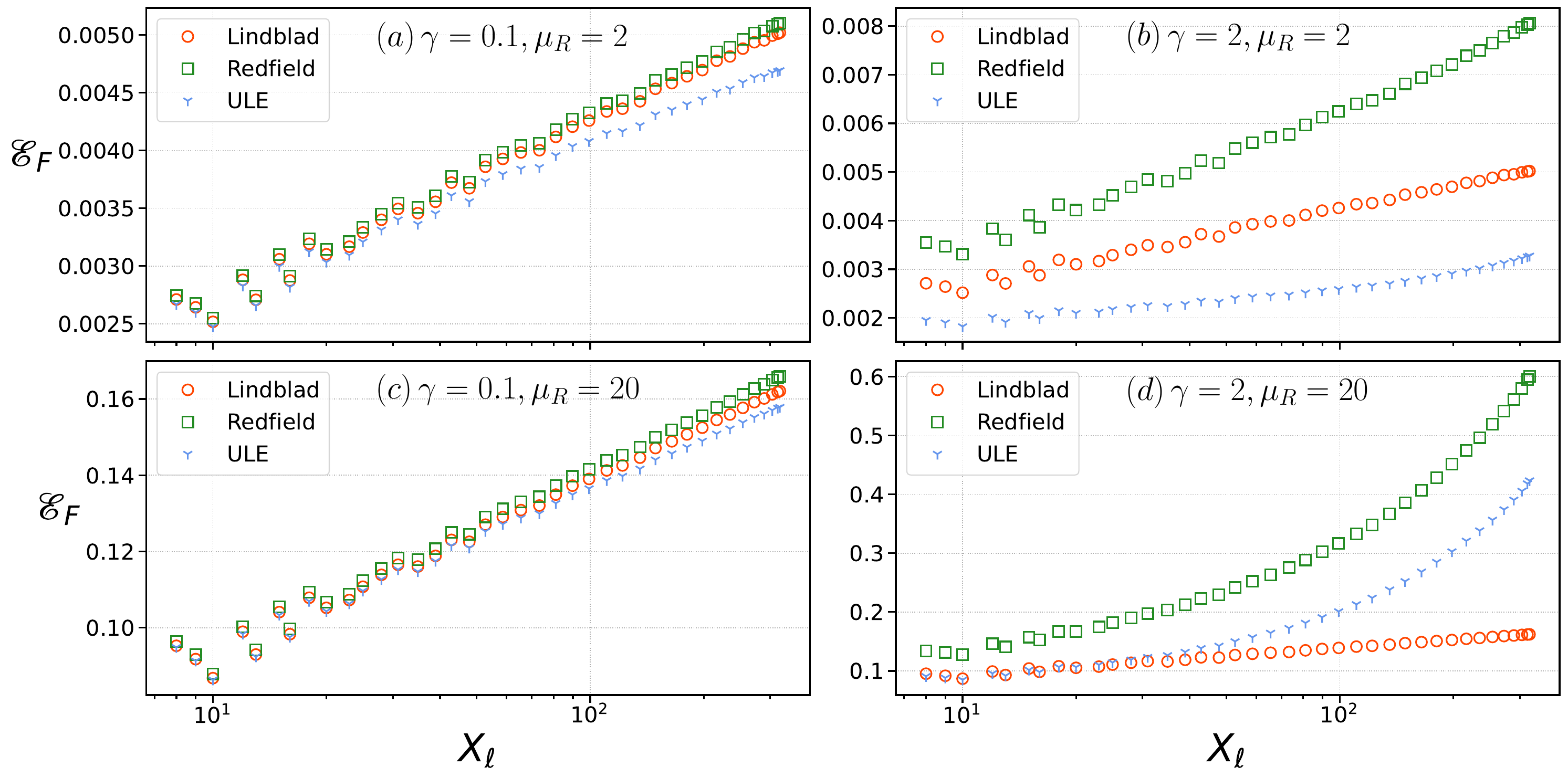}
    \caption{
        Steady-state FLN as a function of the chord length $X_\ell$ for different values of the coupling $\gamma$ and the chemical potential of the right bath $\mu_R$.
        The other parameters are fixed as in Fig.~\ref{fig:example}.
        Panels (a) and (c) show results in the weak-coupling regime for $\gamma=0.1$ and $\mu_R=2$ and $\mu_R=20$, respectively.
	    Panels (b) and (d) show results at strong coupling $\gamma=2$ and $\mu_R=2$ and $\mu_R=20$, respectively.
    }
    \label{fig:deviations}
\end{figure*}

Let us now observe what happens by increasing the value of $\gamma$, thus exiting the weak-coupling regime.
In Fig.~\ref{fig:deviations} we plot the FLN $\mathscr{E}_F$ as a function of the chord length $X_\ell$ for various values of $\gamma$ and $\mu_R$, while keeping the other parameters as in Fig.~\ref{fig:example}.
At weak coupling $\gamma = 0.2$ [panels (a) and (c)], the FLN exhibits a logarithmic increase, as expected, at both $\mu_R = 2$ and $\mu_R = 20$, and the three master equations lead to similar prefactors for the logarithmic growth.
The situation is different at strong coupling $\gamma = 2$ [panels (b) and (d)].
In the case $\mu_R = 2$, we still observe an increase that is compatible with a logarithm, at least for $\ell \lesssim 200$; however, we can notice a slight upward trend at large $\ell$ for the ULE, which suggests a violation of the logarithmic scaling.
This becomes evident if we move to $\mu_R = 20$: the Redfield and ULE approaches clearly yield a superlogarithmic scaling of the FLN at large values of $\ell$.
This contrasts with the results of the nonlocal Lindblad equation, which are compatible with a logarithmic growth at all values of $\gamma$ and $\mu_R$.

\begin{figure}
    \centering
    \includegraphics[width=\columnwidth]{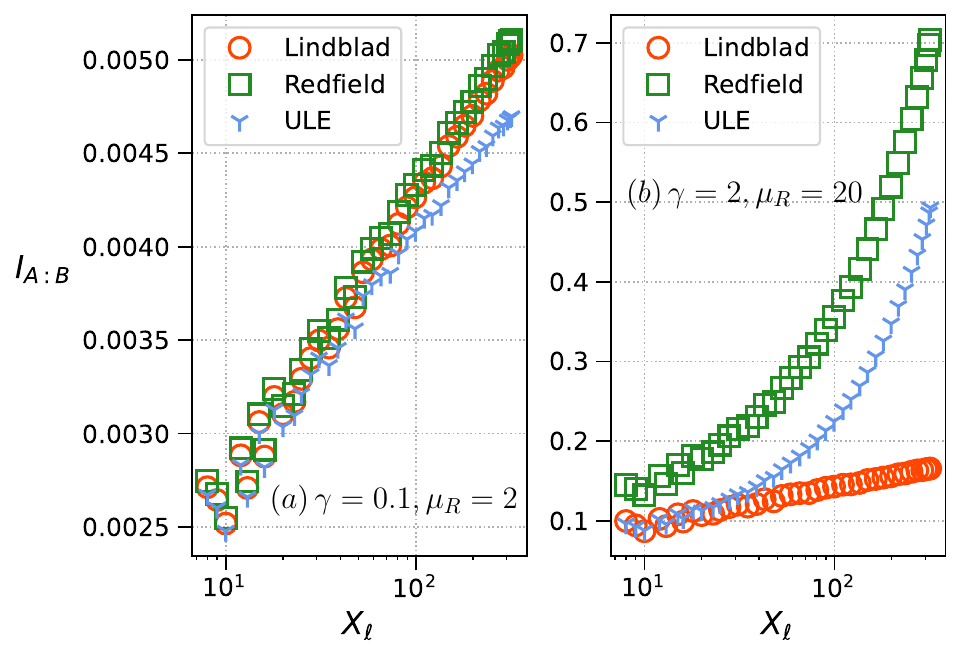}
    \caption{
        Steady-state mutual information as a function of the chord length for the different master equations.
	    The left and right panels show results for $\gamma=0.1$, $\mu_R=2$ and $\gamma=2$, $\mu_R=20$, respectively.
        The other parameters are fixed as in Fig.~\ref{fig:example}.
    }
    \label{fig:mi}
\end{figure}

The presence of a superlogarithmic scaling is confirmed by the behavior of the mutual information $I_{A:B} \coloneqq S_A + S_B - S_{A \cup B}$, where $S_X$ stands for the von Neumann entropy associated with subsystem $X$.
In Fig.~\ref{fig:mi} we plot $I_{A:B}$ as a function of $X_\ell$ for the two cases $\gamma = 0.1$, $\mu_R = 2$ and $\gamma = 2$, $\mu_R = 20$, while the other parameters are fixed as in Fig.~\ref{fig:deviations}.
At weak coupling the data are compatible with a logarithmic scaling, whereas at strong coupling the Redfield and the ULE equations give superlogarithmic scaling, reflecting the behavior of the FLN.
This is surprising, since the mutual information is not a measure of entanglement for mixed states, unlike the FLN.

We finally mention that, while our data exhibit superlogarithmic scaling for the FLN at large values of $\gamma$ and $\mu_R$, it is likely that this effect is present for any value of such parameters, although the crossover length at which it starts to be visible could be pushed at larger and larger $\ell$ upon lowering the coupling strength.
The chain sizes that are accessible within our numerics are not large enough to clarify this, and the same goes for the determination of the precise shape of the superlogarithmic contribution.
Since at weak coupling such effect is not easily seen, it is also possible that it is introduced by the Born-Markov approximation, used to derive both the Redfield and the ULE~\cite{Breuer2002,Nathan2020}.
In fact we should remember that in the strong coupling regime, explored in this section, the master equations described here are not necessarily faithful representations of the true dynamics of the system.


\section{Conclusions} \label{sec:conclusions}

We numerically investigated the steady-state entanglement in a tight-binding critical chain in contact with two thermal reservoirs at its edges.
We showed that logarithmic entanglement scaling, as measured by the FLN, persists beyond the nonlocal Lindblad approach.
Specifically, both the Redfield equation and the ULE yield logarithmic scaling for the steady-state negativity in the regime of weak system-bath coupling, in agreement with the nonlocal Lindblad approach~\cite{DAbbruzzo2022}.
However, upon increasing the coupling strength both approaches give superlogarithmic scaling for the negativity.

Our work opens several directions for future research.
First, it is crucial to characterize the superlogarithmic entanglement scaling contribution.
With the currently accessible system sizes we are not able to clarify whether the superlogarithmic scaling of the negativity follows, e.g., a power law or not.
Moreover, it is important to understand whether such contribution is simply a signal of the breakdown of the Redfield equation and the ULE, or it is a genuine physical effect.

It would be also interesting to perform an analytical characterization of the FLN.
For the nonlocal Lindblad equation, this could be done by applying the results of Ref.~\cite{Eisler2023}, exploiting techniques based on the
Fisher-Hartwig theorem~\cite{Fisher1969,Basor1991,Basor1994}.
However, the solutions of the Lyapunov equations~\eqref{eq:lyapunov} for the Redfield equation and the ULE are not of the Toeplitz form, and a proper framework to extract the asymptotic properties of their spectrum is still missing.

Finally, it would be tempting to understand whether the logarithmic entanglement scaling property survives in the presence of interactions.
While this is a challenging task, it could be explored numerically by employing exact diagonalization methods.


\begin{acknowledgments}
	This study was carried out within the National Centre on HPC, Big Data and Quantum Computing - SPOKE 10 (Quantum Computing) and received funding from the European Union NextGenerationEU - National Recovery and Resilience Plan (NRRP) - MISSION 4 COMPONENT 2, INVESTMENT N. 1.4 - CUP N. I53C22000690001. This work has been supported by the project ``Artificially devised many-body quantum dynamics in low dimensions - ManyQLowD'', funded by the MIUR Progetti di Ricerca di Rilevante Interesse Nazionale (PRIN) Bando 2022 - grant 2022R35ZBF.
\end{acknowledgments}


\appendix

\section{Derivation of the Majorana-Lindblad form for the Redfield equation and the ULE} \label{app:majorana_lindblad}

In this appendix we show how the Redfield equation~\eqref{eq:redfield} and the ULE~\eqref{eq:ULE} can be brought into the Majorana-Lindblad form~\eqref{eq:lindblad}.
We split the derivation in two parts.
First, in Secs.~\ref{app:redfield-m} and~\ref{app:ule-m} we perform the transformation to normal-mode Majorana operators~\eqref{eq:normal_majorana}, realizing that both equations have the same structure;
then, in Sec.~\ref{app:lin-red}, we rearrange the obtained common expression into (pseudo-)Lindblad form.


\subsection{Redfield equation for Majorana operators} \label{app:redfield-m}

We first rewrite Eq.~\eqref{eq:redfield} in the Schr\"{o}dinger picture:
\begin{equation} \label{eq:redfield_schro}
    \frac{d\rho}{dt} + i[H_S, \rho] = \sum_\alpha \int_0^\infty d\tau \, C_\alpha(\tau) [\widetilde{X}_\alpha(-\tau) \rho, X_\alpha] + \text{H.c.}
\end{equation}
Then, we express the operator $X_\alpha$ in Eq.~\eqref{eq:X} using the transformation~\eqref{eq:bogoliubov} to normal modes $d_k$ and the transformation~\eqref{eq:normal_majorana} to corresponding Majorana operators $m_k$:
\begin{equation} \label{eq:X_majo}
	X_\alpha = \sum_{k=1}^N \varphi_{\alpha,k} \qty(d_k + d_k^\dagger) = \sqrt{2}
	\sum_{k=1}^N \varphi_{\alpha,k} m_{2k-1}.
\end{equation}
The quantities $\varphi_{\alpha,k}$ are those defined in Eq.~\eqref{eq:phi}.
The interaction-picture form $\widetilde{X}_\alpha$ of $X_\alpha$ can be obtained through the Baker-Campbell-Hausdorff relation ${e^{-i H_\mathcal{S} t} d_k e^{i H_\mathcal{S} t} = e^{i E_k t} d_k}$.
We obtain
\begin{align}
    \widetilde{X}_\alpha(-\tau) &= \sum_{k=1}^N
    \varphi_{\alpha,k} \qty( e^{iE_k \tau} d_k + e^{-iE_k \tau} d_k^\dagger ) \nonumber \\
    &= \sqrt{2} \sum_{k=1}^N \varphi_{\alpha,k} \qty[ \cos(E_k \tau) m_{2k-1} - \sin(E_k \tau) m_{2k} ]. \label{eq:tildeX_majo}
\end{align}
Putting Eqs.~\eqref{eq:X_majo} and~\eqref{eq:tildeX_majo} into Eq.~\eqref{eq:redfield_schro}, one then easily obtains
\begin{align} \label{eq:redfield_majorana}
    \frac{d\rho}{dt} + i[H_S, \rho] = \sum_{k,q=1}^N
    \big\{ & \mathcal{A}^r_{k q} [m_{2k-1} \rho, m_{2q-1}] \nonumber \\
    + & \mathcal{D}^r_{k q} [m_{2k} \rho, m_{2q-1}] \big\} + \text{H.c.},
\end{align}
where $\mathcal{A}^r_{kq}$ and $\mathcal{D}^r_{kq}$ are defined in Eqs.~\eqref{eq:red_AD}.


\subsection{Universal Lindblad Equation for Majorana operators} \label{app:ule-m}

As for the Redfield equation, we start by reverting the interaction picture in Eq.~\eqref{eq:ULE}.
We find
\begin{multline}
    \frac{d\rho}{dt} + i[H_S, \rho] = \sum_\alpha \int_{-\infty}^\infty d\tau \int_{-\infty}^\infty d\tau' \, \theta(\tau-\tau') \\
    \times g_\alpha(\tau) g_\alpha(-\tau') [\widetilde{X}_\alpha(\tau') \rho, \widetilde{X}_\alpha(\tau)] + \text{H.c.}.
\end{multline}
Using Eq.~\eqref{eq:tildeX_majo} to write $\widetilde{X}_\alpha$ in terms of normal-mode Majorana operators, we obtain
\begin{align} \label{eq:common_majorana_ule}
    \frac{d\rho}{dt} + i[H_S, \rho] = & \sum_{k,q=1}^N \big\{
        \mathcal{A}^u_{k q} [m_{2k-1} \rho, m_{2q-1}] \nonumber \\
        + & \mathcal{B}^u_{k q} [m_{2k} \rho, m_{2q}] + \mathcal{C}^u_{k q} [m_{2k-1} \rho, m_{2q}] \nonumber \\
        + & \mathcal{D}^u_{k q} [m_{2k} \rho, m_{2q-1}]
    \big\} + \text{H.c.},
\end{align}
where
\begin{multline} \label{eq:multi}
    \mathcal{A}^u_{k q} = \sum_{\alpha} \varphi_{\alpha,k} \varphi_{\alpha,q}
    \int_{-\infty}^\infty d\tau \int_{-\infty}^\infty d\tau' \, \theta(\tau-\tau') \\
    \times g_\alpha(\tau) g_\alpha(-\tau') 2 \cos(E_k \tau') \cos(E_q \tau),
\end{multline}
and the other coefficients $\mathcal{B}^u_{kq}$, $\mathcal{C}^u_{kq}$, and $\mathcal{D}^u_{kq}$ are obtained from $\mathcal{A}^u_{kq}$ by replacing $\cos(E_k \tau') \cos(E_q \tau)$ with, respectively, $\sin(E_k \tau') \sin(E_q \tau)$, $\cos(E_k \tau') \sin(E_q \tau)$, and $\sin(E_k \tau') \cos(E_q \tau)$.
Notice the symmetries $[\mathcal{A}^u]^T = \mathcal{A}^u$, $[\mathcal{B}^u]^T = \mathcal{B}^u$, and $[\mathcal{C}^u]^T = -\mathcal{D}^u$.
To recover Eqs.~\eqref{eq:ULE_ABC}, let us introduce the function $\mathcal{I}_{\alpha}(\omega,\omega')$ as
\begin{equation} \label{eq:I}
    \mathcal{I}_\alpha(\omega, \omega') \coloneqq \int_{-\infty}^\infty d\tau d\tau' \, \theta(\tau-\tau') g_\alpha(\tau) g_\alpha(-\tau') e^{i \omega \tau' + i \omega' \tau}.
\end{equation}
By using Eq.~\eqref{eq:I} it is straightforward to obtain Eqs.~\eqref{eq:ULE_ABC} after writing sines and cosines in terms of complex exponentials.

Equation~\eqref{eq:I_better} for $\mathcal{I}_\alpha$, which is more easily numerically computable, is derived as follows (see also Ref.~\cite{Nathan2020}).
First, we employ the relationship
\begin{equation}
    \theta(\tau - \tau') = \frac{1 + \mathrm{sgn}(\tau - \tau')}{2}
\end{equation}
to split $\mathcal{I}_\alpha$ into two contributions as
\begin{equation} \label{eq:split}
    \mathcal{I}_\alpha(\omega,\omega') = \mathcal{I}_\alpha^R(\omega,\omega') + i \, \mathcal{I}^I_\alpha(\omega,\omega').
\end{equation}
For the first one we have
\begin{multline}
    \mathcal{I}_\alpha^R(\omega,\omega') \equiv \frac{1}{2} \int_{-\infty}^\infty d\tau d\tau' \, g_\alpha(\tau) g_\alpha(-\tau') e^{i \omega \tau' + i \omega' \tau} \\
    = \frac{1}{2} (2\pi)^2 \hat{g}_\alpha(-\omega) \hat{g}_\alpha(\omega')
    = \pi \sqrt{ \hat{C}_{\alpha\alpha}(-\omega) \hat{C}_{\alpha\alpha}(\omega') },
\end{multline}
where $\hat{g}_\alpha(\omega) \coloneqq \frac{1}{2\pi} \int_{-\infty}^\infty d\tau \, g_\alpha(\tau) e^{i\omega\tau}=\sqrt{\hat{C}_\alpha(\omega)/2\pi}$ is the Fourier transform of $g_\alpha(\tau)$, and we used Eq.~\eqref{eq:jump}.
For the second term in~\eqref{eq:split} we can now use the fact that $g_\alpha(\tau) = \int_{-\infty}^\infty dE \, \hat{g}_\alpha(E) e^{-iE\tau}$ to write
\begin{multline} \label{eq:Iimag_inter}
    \mathcal{I}_\alpha^I(\omega,\omega') =
    -\frac{i}{2} \int_{-\infty}^\infty d\tau d\tau' \, \mathrm{sgn}(\tau-\tau') \\
    \times g_\alpha(\tau) g_\alpha(-\tau') e^{i\omega\tau' + i\omega'\tau} \\
    = -\frac{i}{2} \int_{-\infty}^\infty dE dE' \, \mathcal{R}(\omega'-E,\omega+E') \hat{g}_\alpha(E) \hat{g}_\alpha(E'),
\end{multline}
where
\begin{multline}
    \mathcal{R}(E,E') \coloneqq \int_{-\infty}^\infty d\tau d\tau' \, \mathrm{sgn}(\tau-\tau') e^{iE\tau + iE'\tau'} \\
    = \int_{-\infty}^\infty d\tau \int_{-\infty}^\tau d\tau' \qty( e^{iE\tau + iE'\tau'} - e^{iE\tau' + iE'\tau} ).
\end{multline}
After the change of variable $\tau' = \tau - s$, we obtain
\begin{equation}
    \mathcal{R}(E,E') = \int_{-\infty}^\infty d\tau e^{i(E+E')\tau} \int_0^\infty d s \qty( e^{-iE's} - e^{-iEs} ).
\end{equation}
Using Eq.~\eqref{eq:sokplem}, this becomes
\begin{equation} \label{eq:last}
	\mathcal{R}(E,E') = -4\pi i \delta(E+E') \mathbb{P}\frac{1}{E}.
\end{equation}
After substituting Eq.~\eqref{eq:last} back into Eq.~\eqref{eq:Iimag_inter}, and after performing the integration over $E'$, we arrive at
\begin{equation}
    \mathcal{I}_\alpha^I(\omega,\omega') = -2\pi \mathbb{P} \int_{-\infty}^\infty \frac{dE}{E} \hat{g}_\alpha(E-\omega) \hat{g}_\alpha(E+\omega'),
\end{equation}
which is the same as Eq.~\eqref{eq:I_better}.


\subsection{Lindblad form for the Redfield equation and the ULE} \label{app:lin-red}

In the previous subsections we showed how the Redfield equation and the ULE can be rewritten using normal-mode Majorana operators in the form
\begin{align} \label{eq:common_majorana}
    \frac{d\rho}{dt} + i[H_S, \rho] = & \sum_{k,q=1}^N \big\{
        \mathcal{A}_{k q} [m_{2k-1} \rho, m_{2q-1}] \nonumber \\
        + & \mathcal{B}_{k q} [m_{2k} \rho, m_{2q}] + \mathcal{C}_{k q} [m_{2k-1} \rho, m_{2q}] \nonumber \\
        + & \mathcal{D}_{k q} [m_{2k} \rho, m_{2q-1}]
    \big\} + \text{H.c.},
\end{align}
where $\mathcal{A} \mapsto \mathcal{A}^r$, $\mathcal{D} \mapsto \mathcal{D}^r$, and $\mathcal{B}, \mathcal{C} \mapsto 0$ for the Redfield case, while $\mathcal{A} \mapsto \mathcal{A}^u$, $\mathcal{B} \mapsto \mathcal{B}^u$, $\mathcal{C} \mapsto \mathcal{C}^u$, and $\mathcal{D} \mapsto -[\mathcal{C}^u]^T$ for the ULE case.

We now derive a master equation in (pseudo-)Lindblad form starting from Eq.~\eqref{eq:common_majorana} with generic coefficients ${\mathcal {A}}_{kq}$, $\mathcal{B}_{kq}$, $\mathcal{C}_{kq}$, and $\mathcal{D}_{kq}$.
First, we write the Hamiltonian as
\begin{equation} \label{eq:ham-re}
    H_S = -\frac{i}{2} \sum_{k,q=1}^N \delta_{kq} E_k \qty( m_{2q} m_{2k-1} - m_{2q-1} m_{2k} ),
\end{equation}
where the presence of $\delta_{kq}$ reflects that the Hamiltonian is translation invariant.
By using Eq.~\eqref{eq:ham-re} in Eq.~\eqref{eq:common_majorana}, one can easily verify that
\begin{equation} \label{eq:lindblad_inter}
    \frac{d\rho}{dt} = \sum_{c,d=1}^{2N} \qty[ V_{cd} m_c \rho m_d - W_{cd} m_d m_c \rho - W^\dagger_{cd} \rho m_d m_c ],
\end{equation}
where $V$ and $W$ are $2N \times 2N$ complex matrices.
The $2 \times 2$ blocks $\mathcal{V}_{kq}$ and $\mathcal{W}_{kq}$ associated with the entries $(2k-1,2q-1)$, $(2k-1,2q)$, $(2k,2q-1)$, $(2k,2q)$ of $V$ and $W$ are respectively given by
\begin{subequations} \label{eq:VW}
    \begin{gather}
        \mathcal{V}_{k q} = \begin{pmatrix}
            \mathcal{A}_{k q} + \mathcal{A}^*_{q k} & \mathcal{C}_{k q} + \mathcal{D}^*_{q k} \\
            \mathcal{D}_{k q} + \mathcal{C}^*_{q k} & \mathcal{B}_{k q} + \mathcal{B}^*_{q k}
        \end{pmatrix}, \label{eq:V} \\
        \mathcal{W}_{k q} = \begin{pmatrix}
            \mathcal{A}_{k q} & \mathcal{C}_{k q} + \delta_{kq} E_k/2 \\
            \mathcal{D}_{k q} - \delta_{kq} E_k/2 & \mathcal{B}_{k q}
        \end{pmatrix}.
    \end{gather}
\end{subequations}
Note that $\mathcal{V}_{kq} = \mathcal{W}_{kq} + \mathcal{W}^\dagger_{qk}$, and therefore one has that $V = W + W^\dagger$.
We now decompose $W$ in its Hermitian and anti-Hermitian components as
\begin{equation}
    M \coloneqq \frac{W + W^\dagger}{2},
    \quad
    L \coloneqq \frac{W - W^\dagger}{2}.
\end{equation}
This allows us to obtain the matrices $L$ and $M$ in Eq.~\eqref{eq:lindblad}.
As it is clear from Eqs.~\eqref{eq:VW}, $M = V/2$.
Moreover, the $2 \times 2$ block $\mathcal{L}_{kq}$ associated to $L$ is
\begin{equation} \label{eq:L}
    \mathcal{L}_{k q} = \frac{1}{2} \begin{pmatrix}
        \mathcal{A}_{k q} - \mathcal{A}^*_{q k} & \delta_{kq} E_k + \mathcal{C}_{k q} - \mathcal{D}^*_{q k} \\
        -\delta_{kq} E_k + \mathcal{D}_{k q} - \mathcal{C}^*_{q k} & \mathcal{B}_{k q} - \mathcal{B}^*_{q k}
    \end{pmatrix}.
\end{equation}
A straightforward computation can now be carried out to show that Eq.~\eqref{eq:lindblad_inter} coincides with Eq.~\eqref{eq:lindblad}.


\bibliography{bibliography.bib}

\end{document}